# Multi-Domain FeFET-Based Pixel for In-Sensor Multiply-and-Accumulate Operations

Md Rahatul Islam Udoy, *Graduate Student Member, IEEE*, Wantong Li, *Member, IEEE,* Kai Ni, *Member, IEEE,* and Ahmedullah Aziz, *Senior Member, IEEE*

*Abstract*— This paper presents an FeFET-based active pixel sensor that performs in-sensor multiply-and-accumulate (MAC) operations by leveraging the multi-domain polarization states of ferroelectric layers. The proposed design integrates a programmable FeFET into a 3-transistor pixel circuit, where the FeFET's non-volatile conductance encodes the weight, and the photodiode voltage drop encodes the input. Their interaction generates an output current proportional to the product, enabling in-pixel analog multiplication. Accumulation is achieved by summing output currents along shared column lines, realizing full MAC functionality within the image sensor array. Extensive HSPICE simulations, using 45 nm CMOS models, validate the operation and confirm the scalability of the design. This compact and power-efficient architecture minimizes data movement, making it ideal for real-time edge computing, neuromorphic vision, and secure sensing applications.

*Index Terms*— edge AI, FeFET, in-sensor computing, in-sensor processing, multiply-and-accumulate.

## I. INTRODUCTION

THE growing demands of real-time edge computing-spanning applications such as autonomous driving, wearable AI, and neuromorphic vision have exposed the inefficiencies of conventional imaging pipelines that separate sensing and computation [1], [2]. In traditional systems, raw sensor data must be transferred to external processing units, incurring significant noise, energy and latency penalties[3]. In-sensor processing offers a promising solution by embedding full or partial computation directly within the sensor array, thereby reducing data movement and enabling faster, more energy-efficient operation[4]. However, simply transplanting processing units from external processors into each pixel sensor can be counterproductive-it increases pixel complexity, and ultimately hampers scalability[5]–[7]. Therefore, a fundamental redesign of the pixel sensor circuit is necessary to ensure that computation is achieved compactly and efficiently.

A core operation in early-stage vision processing is the multiply-and-accumulate (MAC) operation, which underpins the first layer of convolutional neural networks (CNNs)[4], [8]. Performing this layer directly within the sensor is attractive because it operates on raw image data and typically dominates the computational and data bandwidth requirements[2]. In-sensor execution eliminates the need to transfer the entire raw frame off-chip, significantly reducing data movement and energy consumption[8]. Since the first layer primarily extracts low-level features such as edges and textures, it is also well-suited for approximate or analog computation, making it ideal for in-pixel implementations[9]. Additionally, offloading this initial stage improves utilization of downstream accelerators, which are better matched to the higher parallelism and reduced bandwidth demands of later CNN layers[10]. Several prior works have made significant strides in implementing MAC operations at the pixel level, each exploring different trade-offs to balance functionality, precision, and efficiency. Designs based on volatile elements [4], [8] have demonstrated functional viability, but they contribute to higher power consumption. Other approaches achieve non-volatility or enhanced computational capability but often come with a higher transistor count per pixel[1], which can impact scalability and pixel density. These solutions highlight the complexity of integrating analog computing into pixel arrays and underscore the challenge of simultaneously achieving compactness, low power, and non-volatile weight storage. Motivated by these challenges, our work presents a pixel architecture that directly addresses these limitations by combining non-volatility, low power consumption, and minimal transistor count in a compact, scalable design. To demonstrate this, the remainder of the paper is organized as follows: Section II introduces the fundamentals of FeFET devices and pixel circuits. Section III details the proposed MAC-enabled pixel circuit, including its operation and simulation results. Section IV evaluates the area penalty of the compute operation, analyzes the impact of process variations, and benchmarks the proposed design against prior works. Finally, Section V concludes the paper with a summary of key findings and future directions.

## II. MULTI-DOMAIN FeFET & 3-T PIXEL CIRCUIT

A Ferroelectric Field-Effect Transistor (FeFET) is a type of non-volatile transistor that incorporates a ferroelectric capacitor (FeCap) into the gate stack of a traditional MOSFET structure[11]. Fig. 1(a) shows an FeFET structure where $HfO_2$ is used as an FeCap. The key feature of a ferroelectric material is its ability to retain spontaneous polarization even after the removal of an applied electric field and this polarization is called the remnant polarization ($P_R$)[11]. When programmed with a suitable gate voltage pulse, the remnant polarization of the FeCap modulates the channel conductivity[12]. FeFETs offer CMOS compatibility, which make them suitable for

This research was partially funded by NSF grant 2346953.

Md Rahatul Islam Udoy and Ahmedullah Aziz are with the Department of Electrical Engineering and Computer Science, University of Tennessee, Knoxville, TN 37996, USA (email: mudoy@vols.utk.edu; aziz@utk.edu).

Wantong Li is with the Electrical and Computer Engineering Department, University of California, Riverside, CA 92521(email: wantong.li@ucr.edu).

Kai Ni is with the Electrical Engineering Department, University of Notre Dame, IN 46556 (email: kni@nd.edu).

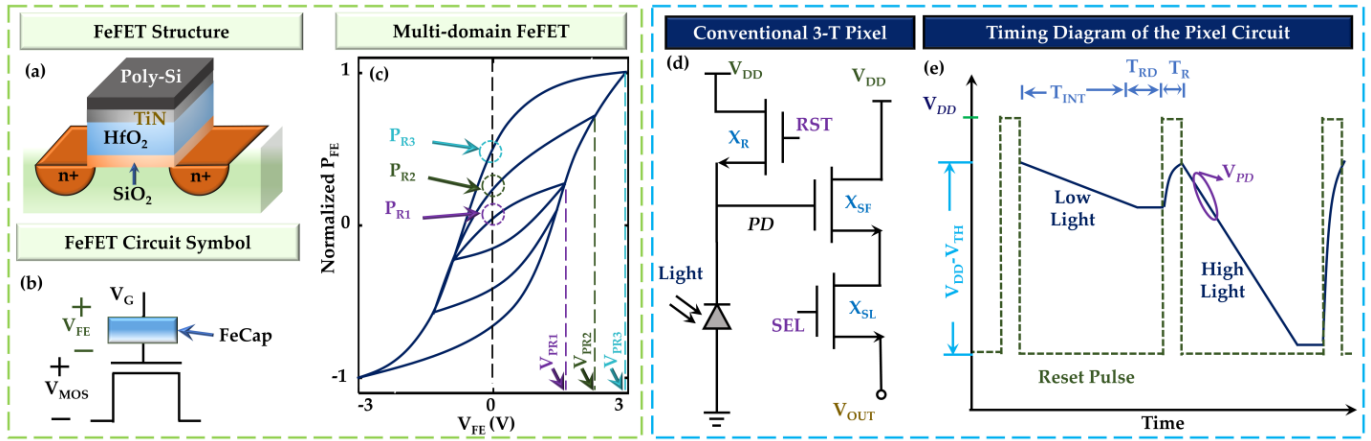

**Fig. 1. (a)** HfO$_2$-based ferroelectric FET (FeFET) structure, and **(b)** its circuit symbol. **(c)** Multi-domain characteristics showing various remnant polarizations (P$_R$) for different programming voltage levels (V$_{PR}$). **(d)** Schematic of a basic 3-transistor (3-T) pixel circuit. Here, X$_R$, X$_{SF}$, and X$_{SL}$ are reset, source-follower, and pixel selector transistors, respectively. **(e)** Timing diagram of a 3-T pixel circuit. A cycle is divided into three phases: T$_{INT}$, T$_{RD}$, and T$_R$, corresponding to the light integration, readout, and reset phases, respectively.

integration in advanced computing architectures[13]. While conventional FeFETs operate with binary polarization states, recent advancements in material engineering have enabled multi-domain behavior within the ferroelectric layer[14]. In multi-domain FeFETs, the polarization is distributed among multiple nanoscale domains that can be partially switched[14]. This allows for multiple stable intermediate polarization levels, resulting in a quasi-analog modulation of the channel conductance[15]. Fig. 1(c) shows the characteristics of such a multidomain FeFET, where for different programming voltages (V$_{PR1}$, V$_{PR2}$, and V$_{PR3}$), different amount of remnant polarizations (P$_{R1}$, P$_{R2}$, and P$_{R3}$) are achieved.

The core component of an image sensor chip is typically a two-dimensional pixel array. Each unit within this array is commonly built using a 3-transistor (3T) pixel circuit, as illustrated in Fig. 1(d). This standard configuration comprises a reverse-biased photodetector along with three transistors: X$_R$, X$_{SF}$, and X$_{SL}$, which function as the reset, source follower, and pixel selector transistors, respectively. The operation sequence for this circuit is depicted in the timing diagram in Fig. 1(e). The process begins with activating the reset transistor (X$_R$), which initializes the photodetector (PD) node voltage (V$_{PD}$) to $V_{DD} - V_{TH,XR}$ (here $V_{TH,XR}$ is the threshold voltage of X$_R$). Alternatively, implementing a PMOS transistor for reset allows the PD node to be fully charged to V$_{DD}$. Following the reset phase, the integration phase starts as the reset transistor is turned off. During this period, the voltage at the PD node gradually drops due to the photocurrent generated by incident light, as described by the equation: $dv/dt = I_{PD}/C_{PD}$. Here, $I_{PD}$ represents the photocurrent, and $C_{PD}$ denotes the intrinsic capacitance of the photodiode. Under higher illumination, $I_{PD}$ increases, leading to a more rapid decline in $V_{PD}$ as illustrated during the second cycle in Fig. 1(e). The transistor X$_{SF}$ operates as a source follower, transmitting the signal from the photodetector node while preserving the accumulated charge. Meanwhile, the third transistor, X$_{SL}$, serves as the pixel selector, enabling the readout of the pixel's signal from the array. The output of this single 3T pixel circuit is denoted by $V_{OUT}$. In a current mode pixel circuit, the output is a current[16].

## III. MULTIPLY-ACCUMULATE PIXEL SENSOR

To enable efficient in-sensor computation for edge intelligence applications, we propose a pixel architecture that integrates computing capabilities directly into the image sensor array. The proposed multiply-accumulate pixel termed MACPix-introduces an FeFET into the pixel design to enable in-sensor analog computation. This modification leverages the multi-domain polarization property of FeFETs. In the pixel circuit, a carefully designed FeFET is connected in series with the source terminal of a PMOS ($X_P$) of the standard 3T pixel circuit (Fig. 2(a)). The remaining components $X_R$ and $X_{SL}$ are reset and pixel selector transistors, respectively. By varying the programming voltage amplitude (PVA) of the FeFET, different levels of remnant polarization ($P_R$) can be achieved, resulting in corresponding variations in its conductance. In this configuration, the FeFET functions as a source degeneration element for the transistor $X_P$, and this pixel generates current ($I_{OUT}$) as an output. We simulate this circuit in HSPICE, which is an industry-grade simulation tool[17]. We use the multi-domain FeCap model from [11]. To transfer the charge value (Q), both the FeCap and the MOSFET models include an additional terminal, as shown in Fig. 2(b). Except for the FeFET ($X_{FE}$), all other transistors in the circuit are modeled using standard 4-terminal MOSFET models. The simulations are carried out using 45 nm high-performance CMOS parameters based on the Predictive Technology Model[18]. The supply voltage is set to 1 V, while the transistors have a channel length of 100 nm and a width of 200 nm.

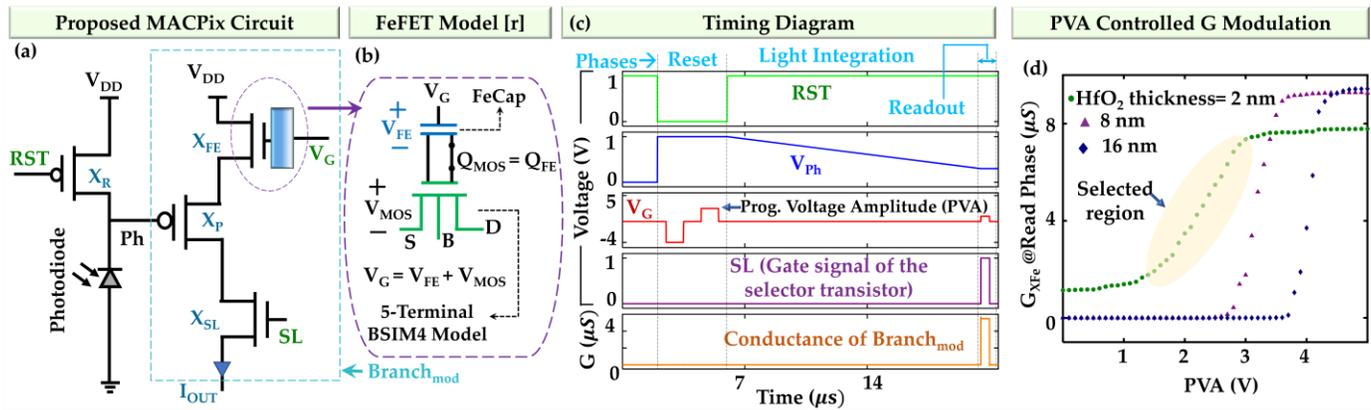

**Fig. 2. (a)** FeFET-based proposed pixel sensor circuit. **(b)** The FeFET is modeled using a modified 5-terminal MOSFET and an HfO$_2$-based FeCap. The extra terminal is used to pass the value of charge. **(c)** The timing diagram of the circuit. FeFET is programmed in the photodiode reset phase. **(d)** Conductance of the FeFET at different programming voltage amplitudes (PVA) and varied HfO$_2$ thickness.

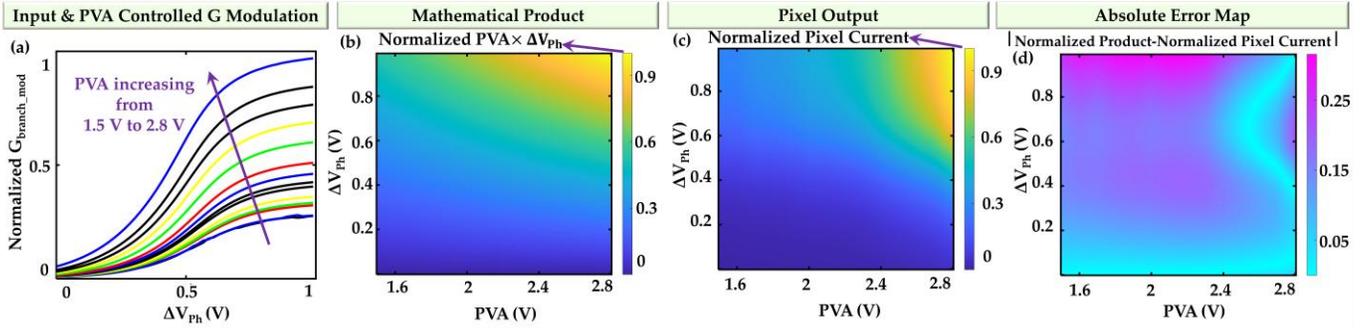

Fig. 3. (a) Conductance modulation by both the PVA and $\Delta V_{Ph}$. The range of PVA (1.5V to 2.8 V) is set by the selected region of Fig. 2(d). Curves shift upward with increasing PVA values. (b) 2D color map of the true mathematical products of varying PVA and $\Delta V_{Ph}$ values. For consistent comparisons, the product values are normalized by their maximum. (c) 2D color map of the pixel currents by exciting the pixel circuit with varying PVA and $\Delta V_{Ph}$ values. (d) The absolute difference between (b) and (c), showing error between true and circuit generated product.

### A. Multiplication

The circuit functionality is shown using a timing diagram in Fig. 2(c). A single cycle can be divided into three phases: reset, light integration, and readout. In the reset phase, the $X_R$ PMOS is turned on and the photodiode node voltage ($V_{Ph}$) reaches $V_{DD}$. In this phase, the programming of the FeFET is also done: at first a negative gate pulse is applied to reset the FeFET to the outer hysteresis loop, then a positive gate pulse is applied to set a certain amount of polarization and this programming voltage amplitude (PVA) determines the amount of polarization and corresponding conductance that will be stored in the FeFET (It is important to note that the negative and positive pulses applied to the gate of the FeFET during the reset phase will not be required in subsequent cycles unless the stored polarization level needs to be changed.). In the light integration phase, the $V_{Ph}$ starts to drop due to photocurrent generation in the photodiode. The amount of voltage drop ($\Delta V_{Ph}=V_{DD}-V_{Ph}$) is proportional to the incident light intensity. At the end of light integration phase, the $X_{SL}$ is turned on and a fixed gate voltage is provided to the FeFET to turn it on. The PVA of the reset phase directly modulates the polarization of the FeFET, thus affecting the conductance of this device. To find the proper region of that modulation, we vary both the thickness of the FeCap and the PVA (in different pixel operation cycles), shown in Fig. 2(d). Based on this analysis, we select a FeCap thickness of 2 nm and a PVA range of 1.5 V to 2.8 V for subsequent simulations. We name the branch of this circuit Branch$_{mod}$, which comprises $X_{FE}$, $X_{SF}$, and $X_{SL}$. The conductance of this branch during the readout phase is named as $G_{branch\_mod}$. Besides PVA, the other parameter that modulates the $G_{branch\_mod}$ is $\Delta V_{Ph}$. The modulation effect of both the PVA and $\Delta V_{Ph}$ is shown in Fig. 3(a). The current output ($I_{OUT}$) of the pixel is linearly proportional to $G_{branch\_mod}$.

In Fig. 3(b), we multiply PVA and $\Delta V_{Ph}$ values, normalize the products, and plot them as a 2D color map. Then we simulate our pixel circuit for all the PVA and $\Delta V_{Ph}$ combinations and calculate the $I_{OUT}$. In Fig. 3(c), we plot another 2D color map to show the normalized calculated currents. Fig. 3(c) approximately matches Fig. 3(b), indicating that the output current of the pixel circuit effectively approximates the result of a multiplication operation between PVA and $\Delta V_{Ph}$. To observe the errors between the true mathematical products and the pixel-generated approximated products (pixel currents), we plot the absolute difference between the normalized mathematical product and the normalized pixel-generated approximated product, as shown in Fig. 3(d). The mean and standard deviation of these errors are 0.13 and 0.07, respectively. We also calculate the Pearson correlation between the true product and the pixel-generated approximated product, which is 0.93, indicating a good linear relationship. The error is higher in the high $\Delta V_{Ph}$ region, which is expected due to the non-linear behavior of the transistors. Accuracy can be improved by discarding higher $\Delta V_{Ph}$ values; however, this comes at the cost of reduced dynamic range in the image sensor. Therefore, a trade-off exists between multiplication accuracy and dynamic range, which can be adjusted based on the specific requirements of the application. Importantly, since the first layer of a CNN primarily extracts low-level features, it is inherently tolerant to such approximations, making it well-suited for analog or in-pixel computation even in the presence of moderate error[9].

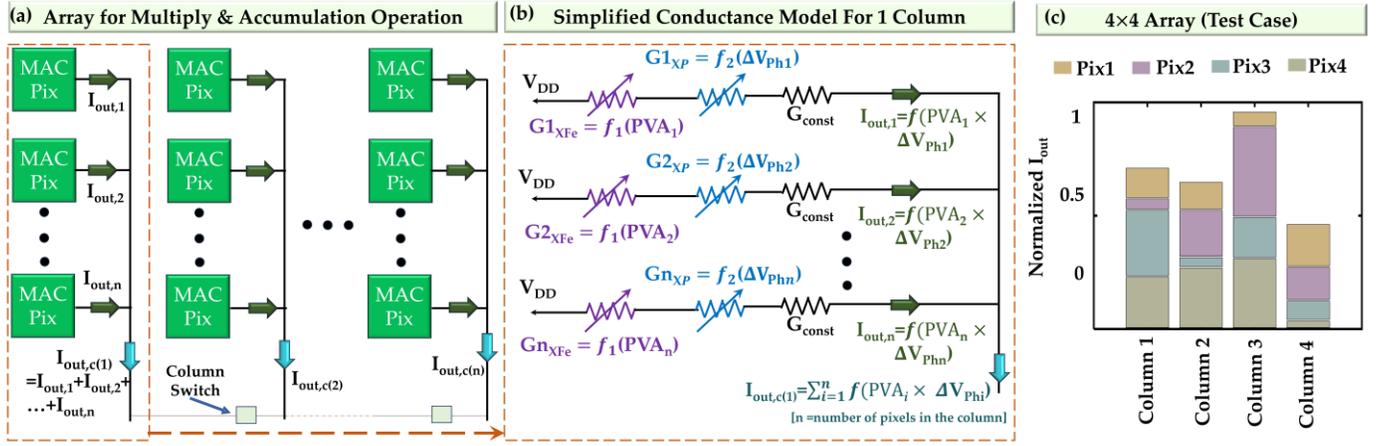

**Fig. 4. (a)** Pixel array configuration where all pixel outputs in a column are connected and their output currents are summed to form a column current ($I_{out,c}$). **(b)** Simplified conductance model, where each line represents the conductance of the Branch$_{mod}$ of the corresponding pixel. **(c)** As a test case, the outputs of 4 pixels in a column are connected and the total current shows the accumulation, and this is verified for three other columns as well. Each pixel's PVA× $\Delta V_{Ph}$ values used for this simulation are shown in Table I.

### B. Accumulation

We make an image sensor array and simulate using our pixel circuits, where the outputs of all the pixels in a column are connected together, as shown in Fig. 4(a). According to Kirchhoff's current law, the column current represents the sum of all the pixel output currents, and thus accumulation is achieved. The simplified conductance model for a column is shown in Fig. 4(b). Here, each line represents the conductance model of the Branch$_{mod}$ of the corresponding pixel, where $G1_{XFE}$ and $G1_{XP}$ are the conductances of the $X_{FE}$ and $X_P$, and $G_{const}$ represents other transistors in the branch. To test the MAC operation, an array is simulated (the PVA×$\Delta V_{Ph}$ values for the pixels are shown in Table I), and the result is plotted in Fig. 4(c), where the total currents of the columns successfully mimic the MAC result.

To use this array in a convolutional neural network, we can use $\Delta V_{Ph}$ as the input and PVA-controlled FeFET conductance as the non-volatile weights. The justification to use $\Delta V_{Ph}$ as the input is: according to the photodiode equation ($\Delta V_{Ph}/\Delta T = I_{PD}/C_{PD}$) this voltage drop is linearly proportional to the photocurrent $I_{PD}$ (where the intrinsic capacitance of the photodiode or $C_{PD}$ is constant, and we keep the light integration time $\Delta T$ fixed). And $I_{PD}$ is also linearly proportional to the photon flux density[19].

TABLE I
PVA× $\Delta V_{Ph}$ VALUES

|  | Col. 1 | Col. 2 | Col. 3 | Col. 4 |
|---|---|---|---|---|
| Pixel 1 | 0.44 | 0.40 | 0.29 | 0.62 |
| Pixel 2 | 0.21 | 0.63 | 1.32 | 0.44 |
| Pixel 3 | 0.97 | 0.19 | 0.6 | 0.28 |
| Pixel 4 | 0.69 | 0.88 | 0.95 | 0.13 |

## IV. AREA PENALTY, PROCESS VARIATION & BENCHMARKING

We compare our MACPix circuit with the regular 3-T pixel circuit to calculate the area penalty per pixel for added MAC functionality. The total pixel area is 4.41 $\mu m^2$. As shown in Table II, for the 45 nm CMOS process, the total area penalty per pixel is 0.072 $\mu m^2$, which accounts for only 1.63% of the total area. This results in an estimated 1.7% fill factor reduction.

TABLE II
PER PIXEL AREA PENALTY

| Contacted Gate Pitch | 160 nm [20] | Area Penalty for using PMOS ($X_P$) | 0.02 $\mu m^2$ |
|---|---|---|---|
| Contacted Diffusion Pitch | 200 nm [20] | Area Penalty for using PMOS ($X_R$) | 0.02 $\mu m^2$ |
| Area Penalty (FeFET) | 0.032 $\mu m^2$ | Total Area Penalty Per Pixel | 0.072 $\mu m^2$ |

To mimic the pixel-to-pixel process variation and noise introduced by the FeFET, we perform a Monte Carlo simulation of 10,000 points, varying different parameters (FeCap thickness, threshold voltage deviation of $X_{FE}$ and $X_P$) according to a Gaussian distribution (as shown in Table III).

TABLE III
MONTE-CARLO VARIATION PARAMETERS

| Parameters | Nominal Value | Variation |
|---|---|---|
| FeCap thickness | 2 nm | ± 9% |
| $\Delta V_{th}$ ($X_{FE}$) | 0 | ± 150 mV |
| $\Delta V_{th}$ ($X_{SF}$) | 0 | ± 150 mV |

The effects of variation on the conductance and remnant polarizations of the FeFET are illustrated in Fig. 5(a) and (b). The statistics are listed in Table IV.

TABLE IV
EFFECT OF VARIATIONS

| Parameters | Values |
|---|---|
| Mean Conductance | 5.355 $\mu S$ |
| Standard Deviation of Conductance | 0.107 $\mu S$ |
| Mean Polarization | 0.89 $\mu C/cm^2$ |
| Standard Deviation of Polarization | 0.053 $\mu C/cm^2$ |

To benchmark our pixel architecture, we present a comparative analysis in Table V, highlighting key parameters from other pixel designs that have been proposed for implementing the first convolutional layer of neural networks. Our pixel design is more compact than other designs. The higher power consumption, compared to [1], is due to the use of current-mode operation in our design.

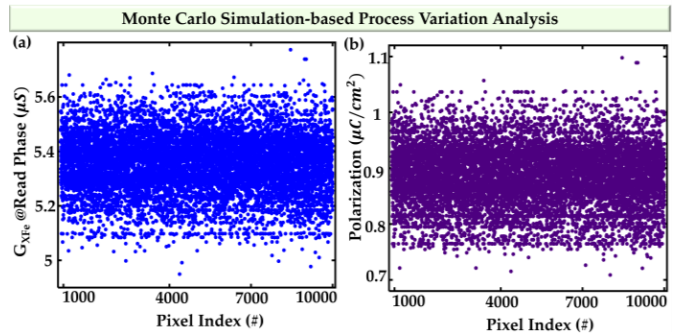

**Fig. 5.** Variation of **(a)** FeFET conductance, and **(b)** polarization due to process variation using Monte-Carlo simulation (variation parameters are shown in Table III). The variation result statistics are shown in Table IV.

TABLE V
BENCHMARKING

|  | Process | $V_{DD}$ (V) | Non-volatile? | Num. of transistors | Pixel pitch ($\mu m$) | Avg. power ($\mu W$) |
|---|---|---|---|---|---|---|
| [4] | 180 nm | 1.8 | No | 3.5T+1 cap | 10 | 1140 |
| [8] | 180 nm | 0.5 | No | 4T | 7.6 | 91 |
| [1] | 45 nm | 1 | Yes | 10 T† | 12.67† | 0.62† |
| This work | 45 nm | 1 | Yes | 4T | 2.1 | 8.33 |

†Normalized estimation per photodiode for fair comparison.

## V. CONCLUSION

This work introduces MACPix, a compact and power-efficient FeFET-based pixel architecture that enables in-sensor multiply-and-accumulate (MAC) operations by leveraging multi-domain polarization states. Simulation results demonstrate that the pixel output current closely approximates the mathematical product of input and weight, achieving a Pearson correlation of 0.93 with a mean error of 0.13 and a standard deviation of 0.07. Full MAC functionality is achieved within the image sensor array using 45 nm process technology, with an area penalty of only 0.072 $\mu m^2$ per pixel, making the design suitable for efficient edge AI applications. Future work will explore material and device fabrication engineering to tailor FeFET characteristics, enabling compensation for MOSFET non-linearity and improving computational accuracy. This work contributes toward enabling compact, low-power, and high-performance vision systems for applications such as autonomous vehicles, smart cameras, and wearable AI.